\documentclass[11pt]{article}
\usepackage{moriond,epsfig}

\def\Journal#1#2#3#4{{#1} {\bf #2}, #3 (#4)}

\def\NPB{{\em Nucl. Phys.} B}
\def\PLB{{\em Phys. Lett.}  B}
\def\PRL{\em Phys. Rev. Lett.}
\def\PRD{{\em Phys. Rev.} D}

\def\JHEP{{\em JHEP} }
\def\EPJ{{\em Eur. Phys. J.} C} 

\def\be{\begin{equation}}
\def\ee{\end{equation}}
\def\bea{\begin{eqnarray}}
\def\eea{\end{eqnarray}}

\begin{document}
\vspace*{4cm}
\title{TOP QUARK PRODUCTION DYNAMICS}

\author{ ERIC LAENEN }

\address{NIKHEF theory group, Kruislaan 409,\\
1098 SJ Amsterdam, The Netherlands}

\maketitle\abstracts{
I discuss various aspects of the dynamics of top quark production 
both via the strong interaction (pair production) and via the
charged current weak interaction (single top production). 
}

\section{The top quark and the Standard Model}
\label{sec:top-quark-standard}

The discovery of the top quark by the CDF and D0 collaborations
at the Fermilab Tevatron in 1995 \cite{topdiscovery} completed,
(together with the discovery \cite{donut} of the $\tau$-neutrino by the DONUT
collaboration, also at Fermilab) the matter sector of the Standard
Model. However, due its unique properties, the importance of the top quark discovery far exceeds
completing a particle table.
First, its large mass, close to the electroweak scale, indicates
that it couples strongly to agents of electroweak symmetry breaking, making
it an interesting probe of that phenomenon. Second, since its large mass 
implies a weak coupling to gluons, it is very suitable for
perturbative QCD precision studies. Third, its large width 
ensures that properties such as its spin are not obscured
by QCD hadronization, but can be studied directly.
Together, these properties make precision study of the top quark and
its quantum numbers, couplings to other particles possible.
An accurate determination of these properties sets
benchmarks for deviations from Standard Model physics.
Because its mass is close to the EW breaking scale,
observables involving the top quark are moreover likely to exhibit effects of new physics first.

The structure of the top quark interactions in the Standard Model is 
quite rich, cf. the color structure 
in the top-gluon interaction 
$g_s [T_a]^{ji}\bar{t}_j \gamma^\mu t_i \, A_\mu^i$, 
the chirality structure, parity violation and flavor mixing in the top-$W$ coupling
$ (g/\sqrt{2})V_{tq} \, \left(\overline{t_L}\gamma^\mu q_L \right) W^+_\mu$,
the parity violation in the top-$Z$ coupling
$  (g/4\cos\theta_W) \, \left(\overline{t}
[1-(8/3)\sin^2\theta_W]\gamma^\mu - \gamma^\mu\gamma^5
 \right) Z_\mu$
and the large strength but simple structure in the top-Higgs Yukawa
interaction $  y_t h \,\overline{t}t$.
Precision studies of these interactions should take place via
observables that have minimal theoretical uncertainty. This requires that
their theoretical descriptions should, if possible, include quantum corrections, 
and that they allow implementation of experimental acceptance restrictions.
This requires fully differential cross sections
$ d^{3n}\sigma/d^3p_1\ldots d^3 p_n$.
Besides obviating the need to extrapolate into, or describe,
regions of phase space where the theory is often less well-behaved, 
and thereby enabling better data-theory comparisons,
such cross sections allow easy construction of different observables
by selective integration over various kinematic variables.

For the foreseeable future top quarks will only be produced at
hadron colliders, both via the strong and the weak interaction.
The study of its production dynamics allows precision measurements
of the strength and structure of the couplings mentioned earlier, 
for verification or falsification.

\section{Top pair production}
\label{sec:top-pair-production}

The largest top quark production cross section involves their
pair production via the strong interaction, and it is this
process that led to the top discovery. A precise theoretical
understanding of pair production dynamics is required to 
match the expected high experimental accuracy, to enable 
precise measurements of various Standard Model properties
and serve as benchmark for new physics signals.
At leading order (LO) this process occurs either via a $q\bar{q}$ or $gg$ initial state. 
The former dominates ($\sim$ 85\%) at the Tevatron, the latter ($\sim$ 90\%)
at the LHC. The NLO corrections are known both for the 
single-particle inclusive  \cite{hqincl} and fully 
differential case \cite{mnr}. Beyond NLO resummations or approximations
based thereupon are available. For the inclusive cross section
the threshold-resummed form for the cross section is, schematically
\begin{equation}
  \label{eq:7}
  \sigma_{q\bar{q}/gg} =  \int_C \frac{dN}{2\pi i} \sigma_0 \, e^{w\, N}\, F_{q\bar{q}/gg}(N,\mu) \,
\exp\left[\ln N g_1(\alpha_s \ln N,\mu) + g_{2,q\bar{q}/gg}(\alpha_s \ln N,\mu) + \ldots  \right]
\end{equation}
with $F$ representing the parton flux, and $\sigma_0$ a hard scattering function.
The approach to threshold is described by $w \rightarrow 0$
(or, equivalently by $N \rightarrow \infty$, where
$N$ is the Mellin (or Laplace) variable conjugate to $w$)
but its definition can vary. A minimal threshold
is $w=s-4m^2$. At fixed $p_T$ one takes $s-4(m^2+p_T^2)$, while if one also fixes rapidity
one may use $w=s+t+u-2m^2$.  
The benefit of expression (\ref{eq:7}) is that it sums large $\ln N$ contributions,
and that it reduces the factorization scale depedence of the hadronic cross section
from  ${\cal O}(\ln N)$ to ${\cal O}(1/N)$. Early resummations 
\cite{hqresumll} for this case varied in their choices and approximations for $w$, $\int_C$, and $g_2$.
The complete next-to-leading resummation \cite{hqresumnll}, requiring inclusion of coherent soft gluon emission \cite{ks},  
based on a minimal choice of contour $C$, found that beyond NLO effects and residual scale dependence
are small. See \cite{sigttrecent} for recent numbers including uncertainties.

Alternatively one may construct a NNLO estimate \cite{klmv} by a two-loop expansion
of the threshold-resummed form of the double-differential cross section \cite{los}, again schematically
\begin{equation}
  \label{eq:1}
    \frac{d^2\sigma_{q\bar{q}/gg}}{dAdB} =  \int_C \frac{dN}{2\pi i} h \, e^{w\, N} \, F_{q\bar{q}/gg}(N,\mu)\,
\exp\left[\ln N g_1(\alpha_s \ln N,\mu) + g_{2,q\bar{q}/gg}(\alpha_s \ln N,A,B) + \ldots  \right]
\end{equation}
where $A,B$ represent kinematics variables: either $t,u$ or $M_{t\bar{t}},\cos\theta_{cm}$. 
The procedure is then to expand (\ref{eq:1}) to NNLO (judging the quality of approximation by comparing to exact NLO
results) and a NNLO estimate after precise matching to NLO. This 
yields all terms of the form $\alpha_s^4\ln^i w, \, i=4,3,2$. Our Tevatron Run II estimate \cite{klmv}
using CTEQ6M parton distributions \cite{cteq6} is
\begin{equation}
  \label{eq:5}
  \sigma_{\bar{t}t}(\mathrm{1.96 \,TeV})  = 7.4 \,\pm\,0.5 \,\pm\, 0.1  \; \mathrm{pb}
\end{equation}
where the first and second error represent the kinematics and scale uncertainty, resp. 
While the inclusive cross section measures essentially the strength of the top
quark QCD coupling, more differential cross sections could serve to explore 
the color structure in this coupling.

\section{Single top production}
\label{sec:single-top-prod}

The top quark may also be produced singly, without its anti-partner (and vice versa), via
the charged current weak interaction.  At lowest order, in a five-flavor scheme, one 
may distinguish production ($s$-channel) via off-shell $W$-bosons that are time-like (e.g. $u\bar{d}\rightarrow t\bar{b}$)
or ($t$-channel) space-like (e.g. $ub\rightarrow td$). Associated production of a $t$ with an on-shell
$W$ is negligible at the Tevatron, but contributes at the LHC. The study of single-top quark 
production dynamics yields a direct measurement of the strength
and the handedness of the top quark charged current coupling, it helps constrain
the bottom-quark density, and leads to new benchmarks for new physics.
The observation of this process at the $5\sigma$ level will require about
$400$pb$^{-1}$ integrated luminosity at the Tevatron. The most promising strategy 
to observe this process was developed in Ref.~\cite{ssw}, featuring in particular
a veto on more than two jets in the central region to reduce the $t\bar{t}$ 
background. The $s$- and $t$-channels can be experimentally defined by 
the number of $b$ tags on the central jets (2 and 1, resp.).

In most studies the single top production mechanism is 
approximated to be independent of its decay. The
quality of this narrow-top-width approximation was
investigated at leading order in Ref.~\cite{helamps} and found to work
quite well.

\begin{figure}
\begin{center}
    \epsfig{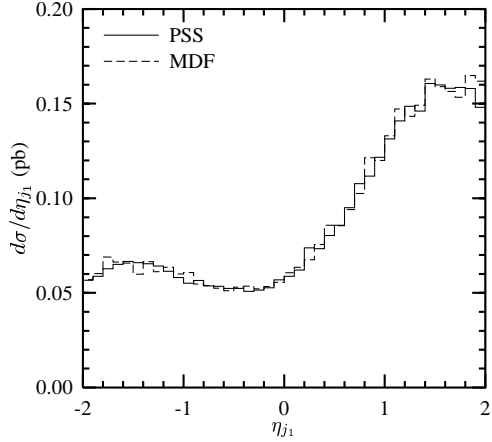}
\end{center}
\caption{NLO pseudorapidity ($\eta_{j1}$) distribution of the highest $p_T$
jet in $t$-channel single top production at the Tevatron
with $\sqrt{S} = 2$ TeV. A $k_T$ cluster algorithm ($\Delta R=1$) was used,
and jets must have $p_T > 20$ GeV and $|\eta | < 2$ to be accepted. Results using the cut-off (PSS) and 
subtraction methods (MDF) agree. \label{fig:one}}
\end{figure}

Of foremost interest will be the direct measurement of the CKM matrix
element $V_{tb}$. Of course, if one assumes the 3-family Standard Model, one need not bother:
$V_{tb}$ is constrained by unitarity to lie in the range $0.999-0.9993$.
Without this assumption, the range is $0.08-0.9993$, so that a direct measurement
of $V_{tb}$ is very revealing. Since the production cross section is proportional to $|V_{tb}|^2$
this quantity can be inferred, using the branching fraction
for $t\rightarrow Wb$ extracted from the pair production signal. 
Thus the eventual accuracy obtained for $V_{tb}$ will depend on, among other things, 
the theoretical accuracy of the single-top production cross sectons.
The pair production cross section was discussed above.
The fully differential NLO single-top production cross sections for
Both $s$ and $t$ channel are given in Ref.\cite{nlo-singletop1}. Both cut-off
\cite{psstwocutoff,pssonecutoff} and dipole subtraction methods \cite{dipole} were used 
to handle intermediate infrared and collinear singularities, and found to give
mutual agreement. Squared matrix elements as well as helicity amplitudes were 
computed to NLO, so that spin information is in principle available. Inclusive
results, in agreement with earlier ones \cite{smithwillenbrock,ssw}, are 
given in table \ref{tab:one},
\begin{table}[t]
\caption{NLO cross sections for single top production ($t+\bar{t}$) at the Tevatron
and LHC for $m_t = 175$ GeV, per channel, and per machine. The indicated error in the third
column is the scale uncertainty. \label{tab:one} }
\vspace{0.4cm}
\begin{center}
\begin{tabular}{|l|l|l|}
\hline
 Channel & $\sqrt{S}$ & $\sigma_{NLO}$(pb)  \\\hline
 t-channel & 1.96 TeV $p\bar{p}$ &  $1.98\pm 0.13 $ \\
           & 14 TeV $pp$ &  $247\pm 12$ \\
 s-channel & 1.96 TeV $p\bar{p}$ &  $0.88\pm 0.09 $ \\
           & 14 TeV $pp$ &  $10\pm 0.9 $\%
\\ \hline
\end{tabular}
\end{center}
\end{table}
for which CTEQ5M1 \cite{cteq5} parton distribution set was used.
An example \cite{nlo-singletop1} of a more differential NLO observable is given in Fig.~\ref{fig:one}.
Finally, I note that optimal bases for verifying the lefthandedness of the top's charged
current coupling were identified in Ref.~\cite{mahlonparke}. A LO analysis including 
backgrounds \cite{ssw} gives good hope for a succesful measurement of this structure.

\end{document}